\documentclass[10pt,a4paper]{article} % 10pt is ignored!
\usepackage{jcappub, natbib}

\allowdisplaybreaks

\usepackage{ifthen}
\usepackage{graphicx}% Include figure files
\usepackage{dcolumn}% Align table columns on decimal point
\usepackage{bm}% bold math
\usepackage{color}
\usepackage{amsmath}
\usepackage{amssymb}
\usepackage{subfigure}

\newcommand{\vp}{\mathbf{p}}
\newcommand{\vq}{\mathbf{q}}
\newcommand{\vs}{\mathbf{s}}
\newcommand{\vx}{\mathbf{x}}
\newcommand{\vu}{\mathbf{u}}

\newcommand{\vk}{\mathbf{k}}
\newcommand{\HH}{\mathcal{H}}
\newcommand{\be}{\begin{equation}}
\newcommand{\ee}{\end{equation}}

\begin{document}

\title{Distribution function approach to redshift space distortions}

\author[a,b,c,d]{Uro{\v s} Seljak,} \emailAdd{useljak@berkeley.edu}
\author[b,e]{Patrick McDonald,} \emailAdd{pvmcdonald@lbl.gov}

\affiliation[a]{Institute of
  Theoretical Physics, University of Zurich, 8057 Zurich, Switzerland}

%\affiliation[b]{Department of Physics, Department of Astronomy and
%Lawrence Berkeley National Laboratory, University of California,
%Berkeley, California 94720, USA}

\affiliation[b]{Lawrence Berkeley National
  Laboratory, University of California, Berkeley, California 94720,
  USA} 

\affiliation[c]{Department of Physics, Department of Astronomy, University of California,
  Berkeley, California 94720, USA} 

\affiliation[d]{Institute for the Early Universe, Ewha Womans
  University, Seoul 120-750, S. Korea}

%\affiliation[d]{Canadian Institute for Theoretical Astrophysics, 60
%St. George Street Toronto, Ontario, M5S 3H8, Canada}

\affiliation[e]{Physics Dept., Brookhaven National Laboratory,
  Building 510A, Upton, NY 11973-5000, USA}
\abstract{
We develop a phase space distribution function approach to redshift space 
distortions (RSD), in which the redshift space density can be written as a sum 
over velocity moments of the distribution function. These moments are density 
weighted and have well defined physical interpretation: their lowest orders are
density, momentum density,  and stress energy density. The series expansion is 
convergent if $k\mu u/aH<1$, where $k$ is the wavevector, $H$ the Hubble 
parameter, $u$ the typical gravitational velocity and $\mu=\cos \theta$, with
$\theta$ being the angle between the Fourier mode and the line of sight. We 
perform an expansion of these velocity moments into helicity modes, which are 
eigenmodes under rotation around the axis of Fourier mode direction, 
generalizing the scalar, vector, tensor decomposition of perturbations to an 
arbitrary order. We show that only equal helicity moments correlate and derive 
the angular dependence of the individual contributions to the redshift space 
power spectrum. We show that the dominant term of $\mu^2$ dependence on large 
scales is the cross-correlation between the density and scalar part of momentum
density, which can be related to the time derivative of the matter power 
spectrum. Additional terms contributing to $\mu^2$ and dominating on small 
scales are the vector part of momentum density-momentum density correlations, 
the energy density-density correlations, and the scalar part of anisotropic 
stress density-density correlations. The second term is what is usually 
associated with the small scale Fingers-of-God damping and always suppresses 
power, but the first term comes with the opposite sign and always adds power. 
Similarly, we identify 7 terms contributing to $\mu^4$ dependence. Some of the 
advantages of the distribution function approach are that the series expansion 
converges on large scales and remains valid in multi-stream situations. 
We finish with a brief discussion of implications for RSD in galaxies
relative to dark matter, highlighting the issue of scale dependent bias of 
velocity moments correlators. 
}
%\pacs{}
\arxivnumber{11XX.XXXX}

\maketitle

%%%%%%%%%%%%%%%%%%%%%%%%%%%%%%%%%%%%%%%%%%%%%%
% Section 1 Introduction
%%%%%%%%%%%%%%%%%%%%%%%%%%%%%%%%%%%%%%%%%%%%%%
\section{Introduction}\label{sec:intro}

Galaxy clustering has traditionally been one of the most important ways to extract 
cosmological information. Galaxies are 
not a faithful tracer of dark matter, as their clustering strength is biased relative to the dark matter. 
However, they are expected to follow the same gravitational 
potential as the dark matter and hence have the same velocities. This is not observable through 
angular clustering, which is only sensitive to correlations transverse to the line of sight. 
It is however detectable in redshift surveys, because the redshift of the galaxy does not 
provide information only on the radial distance, but also on the radial velocity through the Doppler shift. 
This induces anisotropies in the clustering, which are generically called redshift space distortions (RSD) 
\cite{1998ASSL..231..185H}. 
They provide an opportunity to extract information on the dark matter clustering directly. 
On large scales clustering of galaxies along the line of sight is enhanced
relative to the transverse direction due to peculiar motions and this allows
one to determine the ratio of logarithmic rate of growth $f$ to bias $b$ \cite{1987MNRAS.227....1K}. Combining 
the statistics from different lines of sight 
one can eliminate the unknown bias and measure 
directly the logarithmic rate of growth times the amplitude. 

It has been argued that using RSD information could greatly increase our knowledge of cosmological 
models, including tests of dark energy and general relativity \cite{2009MNRAS.397.1348W,2009JCAP...10..007M,2011arXiv1104.3862B}. 
Galaxy clustering has clear 
advantages over the alternatives such as weak lensing: it is intrinsically 3-dimensional, thus providing 
better statistics, and it has high signal to noise. 
While most of the predictions in the literature are model dependent, a generic statement can be made that if systematic
effects were perfectly understood RSD would be one of the most powerful techniques for such studies. 
The main problem with RSD is that nonlinear velocity effects extend to rather large scales and 
give rise to a scale dependent and angular dependent clustering signal. It is easy to see these effects in any 
real redshift survey: one sees elongated features along the line of sight, 
called the fingers-of-god (FoG) effect, 
which are caused by random velocities inside virialized objects such as 
clusters, which scatter galaxies 
along the radial direction in redshift space, even if they have a localized 
spatial position in real space. 
This is just an extreme example and other related effects, such as nonlinear infall streaming 
motions, also cause nonlinear corrections. 
This means that one needs to understand these and separate them from the nonlinear evolution of the dark 
matter and from the nonlinear relation between the galaxies and the dark matter, both of which also give rise to 
a scale dependent bias \cite{1998MNRAS.301..797H}. 

Several recent studies have investigated these nonlinear effects 
\cite{2004PhRvD..70h3007S,2007MNRAS.374..477T,2010PhRvD..82f3522T,2011MNRAS.410.2081J,2011arXiv1103.3614T,2011arXiv1105.4165R,2011arXiv1105.5007S}, 
some limiting the analysis 
to dark matter only and some also including galaxies or halos. The common denominator of these studies is 
that they are based on various ansatzes for the scale and angular dependence of RSD, typically 
combined with some perturbation theory analysis. This has the advantage of having just a few free parameters, so 
that if the ansatz is accurate one can model the effects accurately. The reverse is also true and the problem is 
that it is difficult to make general statements regarding the range of validity for any given model. 
There is another problem connected to perturbation theory, in that the usual perturbation theory makes a single stream
approximation, which we know breaks down on small scales inside the virialized halos 
(indeed, FoG are a manifestation of multi-streaming on small scales).  

In this paper we 
present a different approach to RSD: we use a distribution function approach to 
show that one can make a series expansion of RSD, which is convergent on 
sufficiently large scales 
and we derive the most general form of RSD correlator allowed by the
symmetries. In this paper we present the formal derivations and conceptual implications, 
reserving all the applications to future work. 
The structure of this paper is as follows: in section 2 we develop the distribution function approach
to RSD and derive the helicity decomposition. In section 3 we discuss the power spectra, and use rotation symmetries to derive the most general 
form of the RSD correlator. We also discuss the lowest order contributions
and connect them to physical quantities such as density, momentum, stress energy tensor etc. 
This is followed by a discussion in section 4. 

%%%%%%%%%%%%%%%%%%%%%%%%%%%%%%%%%%%%%%%%%%%%%%
% Section 2 RSD
%%%%%%%%%%%%%%%%%%%%%%%%%%%%%%%%%%%%%%%%%%%%%%

\section{Redshift-space distortions from the distribution function}
\label{sec:theory}
The exact evolution of collisionless particles is described by the
Vlasov equation \cite{1980lssu.book.....P}.  Following the discussion by
\cite{2011JCAP...04..032M}, we start from the distribution function of
particles $f(\vx,\vq,\tau)$ at a phase-space position $(\vx,\vq)$ and 
at conformal time $\tau$ in order
to derive the perturbative redshift-space distortions.  Here $\vx$ is the 
comoving position and $\vq=\vp/a=m\vu$ is the comoving momentum, where 
$\vu=d\vx/d\tau$. 
In the following we will omit the time dependence, i.e we will 
write $f(\vx,\vq)$. 
The density
field in real space is obtained by averaging the distribution function
over momentum:
\begin{equation}
  \rho\left(\vx\right)\equiv m_p \int d^3 \vq ~f\left(\vx,\vq\right),
\end{equation}
where $m_p$ is the particle mass
and $a=1/(1+z)$ is the scale factor ($z$ is the redshift).
In redshift space the position is distorted by peculiar
velocities, thus the comoving redshift-space coordinate for a particle is given
by $\vs=\vx+\hat{r}~u_\parallel/ \HH$, where $\hat{r}$ is
the unit vector pointing along the observer's line of sight, $u_\parallel$ is 
the radial velocity, 
$m_pu_\parallel=q_\parallel = \vq\cdot \hat{r}$, and $\HH=aH$, where $H$ is the Hubble
parameter.  Then the mass density in redshift space is given by
\begin{equation}
  \rho_s\left(\vs\right)=
  m_p~ \int d^3\vx~ d^3\vq~ f\left(\vx,\vq\right)
  \delta^D\left(\vs-\vx-\hat{r}\frac{u_\parallel}{  \HH}\right)=
  m_p~ \int d^3\vq~ f\left(\vs-\hat{r}
  \frac{u_\parallel}{  \HH},
  \vq\right)~. \label{eq:rhos}
\end{equation}
By Fourier transforming equation \ref{eq:rhos}, we find
\begin{eqnarray}
  \rho_s\left(\vk\right)&=&
  m_p~ \int d^3\vx~ d^3\vq~ f\left(\vx,\vq\right)
  e^{i \vk\cdot \vx + i k_\parallel u_\parallel/  
  \HH} \nonumber \\ &=&
  m_p~ \int d^3\vx~ e^{i \vk\cdot \vx } 
  \int d^3\vq~ f\left(\vx,\vq\right)
  e^{i k_\parallel u_\parallel/  \HH} ~,  \label{eq:rhok}
\end{eqnarray}
where $\vk$ is the wavevector in redshift space, corresponding to the
redshift-space coordinate $\vs$.

Now we expand the second integral in equation \ref{eq:rhok} as a Taylor series 
in $k_\parallel u_\parallel/  \HH$,
\begin{eqnarray}
  m_p~ \int d^3\vq~ f\left(\vx,\vq\right)
  e^{i k_\parallel u_\parallel/  \HH} 
  &=&m_p~ \int d^3\vq~ f\left(\vx,\vq\right)
  \sum_{L=0}\frac{1}{L!} \left(i k_\parallel u_\parallel/  \HH\right)^L
  \nonumber \\&=&\bar{\rho}\left[\sum_{L=0}\frac{1}{L!}
    \left(\frac{i k_\parallel}{\HH}\right)^L T_\parallel^L(\vx)\right] ~,
\end{eqnarray}
where $\bar{\rho}$ is the mean mass density and
\begin{equation}
  T_\parallel^L(\vx)={m_p \over \bar{\rho}} 
  ~ \int d^3\vq~ f\left(\vx,\vq\right) 
  u_\parallel^L
%= \left\langle \left(1+\delta(\vx)\right) u_\parallel^L(\vx) \right\rangle_{\vx}
, \label{eq:q_def}
\end{equation}
where in the last expression the integral over phase space assures that the quantity 
is defined in terms of a
sum over all particles at position $\vx$. 
For a single stream at $\vx$ this is just 
$(1+\delta(\vx)) u_\parallel^L(\vx)$.
These
are thus  radial components of the moments of the distribution function and
the distribution function description allows for inclusion of both bulk velocities and
multi-streamed velocities.  
Note that these quantities are
mass-weighted, and so well-defined for any system: one just sums 
over all the particles in the system weighting each one by the appropriate 
power of their radial velocities. If the field needs to be defined on a grid 
a simple assignment scheme of particles to the grid 
suffices, and empty grid cells are assigned a value of 0. 
We note this to contrast it with volume weighted quantities which need to 
be 
defined even if there are no particles assigned to a given grid cell, which is 
often impossible 
for sparse biased tracers, specially in underdense regions. We return to this issue later. 

 The Fourier component of the
density fluctuation in redshift space is
\begin{equation}
  \delta_s(\vk)= 
  \sum_{L=0}\frac{1}{L!}
  \left(\frac{i k_\parallel}{\HH}\right)^L T_\parallel^L(\vk) ~,
  \label{eq:deltak}
\end{equation} 
where $T_\parallel^L(\vk)$ is the Fourier transform of $T_\parallel^L(\vx)$.
\be
T_\parallel^L(\vk)=\int d^3\vx~ T_\parallel^L(\vx)
  e^{i \vk\cdot \vx}.
\ee
For $L=0$ we have
$T_\parallel^0(\vk)=\delta(\vk)$, 
the density fluctuation in real space. 

\subsection{Angular decomposition of moments of distribution function}

The objects $T_\parallel^L(\vx)$ introduced in equation \ref{eq:q_def}
are radial components of moments of the
the distribution function, which are rank $L$ tensors,
\be
T^L_{i_1,i_2,..i_L}={m_p \over \bar{\rho}}
  ~ \int d^3\vq~ f\left(\vx,\vq\right)u_{i_1}u_{i_2}...u_{i_L}.
\label{eqTfulldef}
\ee
The real-space density field corresponds to
$L=0$, i.e. zeroth moment, 
the $L=1$ moment corresponds to the momentum density, 
$L=2$ gives the stress energy density tensor etc.  
These objects are symmetric under exchange of any two indices and 
have $(L+1)(L+2)/2$ independent components. They can be decomposed into 
helicity eigenstates under rotation around $\vk$, as we do next. 

Since translational symmetry guarantees that each Fourier mode is only correlated 
with itself, we can work with each Fourier mode separately, and add them  appropriately
in the end when we discuss the power spectra. By symmetry we may take  $\vk$ to be along $z$-axis. 
We can decompose the distribution function into spherical harmonics, 
\be
f(\vk,q,\theta,\phi)=\sum_{l=0}^{\infty}\sum_{m=-l}^{m=l}f_l^m(\vk,q)Y_{lm}(\theta,\phi),
\label{flm}
\ee
where $q$ is the amplitude of the momentum (often a term $(-i)^{l}\sqrt{{4 \pi \over 2l+1}}$ is inserted into this 
expansion, but we will drop all such terms here). 
The components $f_l^m(\vk,q)$ are helicity eigenmodes (i.e., eigenmodes of angular momentum component in $z$-direction 
$L_z=-i\partial/\partial \phi$) and under rotation by angle $\psi$ around the $z$-axis 
they transform as 
\be
f_l^m(\vk,q)'=e^{im\psi}f_l^m(\vk,q).
\ee
This follows from the transformation properties of spherical harmonics. 
A quantity which transforms under rotation according to this equation is said to 
have helicity $m$. A quantity with helicity $0$ is called a scalar, that with helicity 
$m=\pm 1$ is called a vector and that with $m=\pm 2$ a tensor, but the expansion goes to 
arbitrary values of $m$. 

Moments of the distribution function are defined in terms of integrals of velocity moments 
over the distribution function. 
We can define helicity eigenstates of moments of the distribution function as 
\be
T_l^{L,m}(\vk)={4\pi m_p \over \bar{\rho}}~\int q^2dq u^L f_l^m(\vk,q).
\label{tllm}
\ee
Note that each term $T_{l}^{L,m}$ contains $L$ powers of velocity 
$u$ ($u\equiv\left|\vu\right|$, and recall that $\vq = m \vu$). 

A general rank $L$ tensor $T^L_{i_1,i_2,..i_L}$ can be decomposed into 
$2L+1$ helicity eigenmodes $T_L^{L,m}$ ($m=-L,..,L$) and 
additional components  
formed 
by a product of a scalar $u^2$ with rank $L-2$ tensors. This gives 
additional $2(L-2)+1$ 
helicity eigenmodes 
$T_{L-2}^{L,m}$ ($m=-(L-2)..(L-2)$), and additional components formed 
again by $u^2$ and rank $L-4$ tensor, which gives additional
$2(L-4)+1$ helicity eigenmodes
$T_{L-4}^{L,m}$ ($m=-(L-4)..(L-4)$) etc. 

For the lowest terms we have
rank 1 tensor $T_i^1$, momentum density, which is a 3-vector in the usual geometrical context, 
and can be decomposed into a $m=0$ helicity scalar component $T_1^{1,0}$ 
and two $m=\pm 1$ helicity vector components $T_1^{1,\pm 1}$. 
A general symmetric rank 2 tensor $T_{ij}^2$ has 6 independent components. These 
can be decomposed into an isotropic rank 0 helicity scalar term $T^{2,0}_0=(1+\delta)u^2$, 
which corresponds to the energy density, and 5 $l=2$ components: one helicity scalar part of anisotropic 
stress tensor $T^{2,0}_2$, two helicity vector part of anisotropic stress tensor $T^{2,\pm 1}_2$
and two helicity tensor part of anisotropic stress tensor $T^{2,\pm 2}_2$.
At $L=3$ we have 10 independent components, of which 7 are $T^{3,m}_3$, i.e. $l=3$, $m=-3,..,3$, 
and 3 are $T^{3,m}_1$, $l=1$, $m=-1,0,1$ tensors, 
formed by taking isotropic $u^2$ and multiplying it with a 3-vector $u_i$, the latter of which can be 
decomposed into $m=-1,0,1$ components. 

One can show that $2L+1+2(L-2)+1+2(L-4)+1+...=(L+1)(L+2)/2$, so this decomposition 
gives the required number of independent components of a general symmetric tensor of rank $L$. 
In analysis of 
general relativity it is customary to expand the metric and stress energy tensor into scalar ($m=0$), vector ($m=\pm 1$)
and tensor ($m=\pm 2$) helicity modes (SVT decomposition). No higher order helicity modes 
are needed, since only tensors of rank 0, 1 and 2 enter 
into the description of the metric and energy momentum tensor. 
In contrast the moments of distribution function contain tensors of arbitrary rank and the 
expansion in equations \ref{flm},\ref{tllm} is the appropriate generalization of the SVT decomposition.

So far we worked in the basis defined by $\vk$ pointing in $z$ direction. 
In general we are interested in computing the components of the moments in the radial direction $\hat{r}$. 
If $\hat{r}$ is parallel to $\vk$ then only $m=0$ components contribute, while for a general direction all of them do.
The angular dependence of the moments is obtained
by performing a rotation of the basis from $z||\vk$ to $z'||\hat{r}$. 
We can achieve this by rotating by $\phi$ around $z$ and then by $\theta$ around the axis perpendicular to $z$, $z'$, 
so in terms of the general rotation by 3 Euler angles we have $T_{l}^{L,m'}=\sum_{-m}^{m}D^l_{m,m'}(\phi,\theta,0)T_{l}^{L,m}$, 
where $D^l_{m,m'}(\phi,\theta,\phi')$ is the general rotation matrix of spin $l$ associated with the 3 Euler angles $\phi$, $\theta$, $\phi'$ 
(we do not need to perform the rotation around $z'$ by $\phi'$). 
Since $u_{\parallel}$ is invariant under rotation around $z'||\hat{r}$ only $m'=0$ survive. 
The rotation matrix is given by the spherical harmonics, $D^l_{0,m}(\phi,\theta,0)=\sqrt{4\pi/(2l+1)}Y_{lm}(\theta,\phi)$. 
Combining all together we find
\be
T_{\parallel}^L(\vk)=\sum_{(l=L,L-2,..)}\sum_{m=-l}^{m=l}n_l^LT_l^{L,m}(\vk)Y_{lm}(\theta,\phi),
\label{tpl}
\ee
where $N_l^L$ is a constant independent of angle whose numerical value  will not be needed.

\section{Power spectra}\label{sec:power_th}
We will adopt a plane-parallel approximation, where   
only the angle between the line of sight and the Fourier mode 
needs to be specified. 
The redshift-space power spectrum is defined as $\langle \delta_s(\vk)\delta_s^*(\vk')\rangle=P^{ss}(\vk)\delta_D(\vk-\vk')$.
Equation \ref{eq:deltak} gives,
\begin{equation}
  P^{ss}(\vk)=
  \sum_{L=0}^{\infty}\sum_{L'=0}^{\infty}\frac{\left(-1\right)^{L'}}{L!~L'!}
  \left(\frac{i k_\parallel}{\HH}\right)^{L+L'} P_{LL'}(\vk) \label{eq:p_ss0} ~,
\end{equation}
where $P_{LL'}(\vk)\delta(\vk-\vk')=\langle T^{L}_\parallel(\vk) (T^{*L'}_\parallel(\vk') \rangle$.  
Note that $P_{LL'}(\vk)=P_{L'L}(\vk)^*$ so that the
total result is real valued, as expected. Thus we only need to consider the terms $P_{LL'}(\vk)$ with 
$L\le L'$, each of which comes with a factor of 2 if $L \ne L'$ and 1 if $L=L'$. 
We can also write $k_{||}/k=\cos \theta=\mu$, 
\begin{equation}
  P^{ss}(\vk)=\sum_{L=0}^{\infty}\frac{1}{L!^2}\left(\frac{ k\mu}{\HH}\right)^{2L} P_{LL}(\vk) +
  2\sum_{L=0}^{\infty}\sum_{L'>L}\frac{\left(-1\right)^{L'}}{L!~L'!}
  \left(\frac{i k\mu}{\HH}\right)^{L+L'} P_{LL'}(\vk) \label{eq:p_ss} ~.
\end{equation}

Next we want to insert the helicity decomposition of equation \ref{tpl} and 
consider the implications of rotational symmetry on the power spectrum. 
Each term $P_{LL'}(\vk)$ contains products of multipole moments 
\be
T_l^{L,m}Y_{lm}(\theta,\phi)[T_{l'}^{L',m'}Y_{l'm'}(\theta,\phi)]^* \propto e^{i(m-m')\phi}.
\ee
Upon averaging over the azimuthal angle $\phi$ of Fourier modes all the terms 
with $m\ne m'$ vanish. Another way to state this is that upon rotation by angle $\Psi$ the correlators
pick up a term $e^{i(m-m')\Psi}$, and in order for the power spectrum to be rotationally invariant we 
require $m=m'$. Putting it all together we find
\be
P_{LL'}(\vk)=\sum_{(l=L,L-2,..)}\sum_{(l'=L',L'-2,..;\; l'\ge l)}\sum_{m=0}^{l}P^{L,L',m}_{l,l'}(k)P_l^m(\mu)P_{l'}^m(\mu),
\label{pll}
\ee
where $P_l^m(\mu=\cos \theta)$ are the associated Legendre polynomials, which determine the $\theta$ angular dependence
of the spherical harmonics, $Y_{lm}(\theta,\phi)=\sqrt{(2l+1)(l-m)!/4\pi(l+m)!}P_l^m(\cos \theta) e^{im\phi}$.
We 
absorbed all of the terms that depend on $l$ and $m$ and various constants into the definition of power spectra $P^{L,L',m}_{l,l'}(k)$, 
replaced the two helicity states $\pm m$ by a single one with $m>0$, since their $\theta$ angular dependencies are the same, 
and we absorbed the factor of 2 into the definition of $P^{L,L',m}_{l,l'}(k)$. We also require $l'\ge l$ and 
absorb the factor of 2 into the definition of $P^{L,L',m}_{l,l'}(k)$ since the two terms have the same
angular structure. 
Note that due to statistical isotropy the spectra $P^{L,L',m}_{l,l'}(k)$ depend only on amplitude of $k$, 
i.e. we have 
\be
P^{L,L',m}_{l,l'}(k) \propto \langle T_l^{L,m}(\vk)(T_{l'}^{L',m}(\vk))^* \rangle .
\ee
All the angular structure is thus in associate Legendre polynomials 
$P_l^m(\mu)$. 

Equations \ref{eq:p_ss} and \ref{pll} are the main result of this paper. 
They show that there exists a well defined expansion in terms of cross and auto-power spectra of velocity 
moments. The expansion parameter is roughly defined as $k\mu u/\HH$, where 
$u$ is related to a typical gravitational velocity of the system (which should be of the order of hundreds of km/s, 
but note that we take higher and higher powers of these velocities in the series). 
The expansion is convergent if the expansion parameter is less than unity. 

In terms of perturbation theory there is a close, but not one to one, relation between the lowest order of perturbation 
theory and the order of the moment expansion. 
Assuming $\delta$ and $ku/\HH$ are of the same order,  
the lowest order of the contribution in terms of powers of power spectrum (i.e., quadratic in $\delta$) is  
$(L+L')/2$ if 
$L+L'$ is even and $L>0$, and $(L+L'+1)/2$ if odd and $L>0$, while for $L=0$ it is $L'/2+1$ if $L'$ even 
or $(L'+1)/2$ if $L'$ odd, but of course all 
higher order terms also enter. 

These equations also show that there is a close relation between the order of the moments and their 
angular dependence. To understand the angular dependence we first note that 
associated Legendre polynomial $P_l^m(\mu)$ contains powers from 1 to $\mu^{l-m}$ for even $l$, and from
$\mu$ to $\mu^{l-m}$, for odd $l$, 
and is always multiplied with a power of $(1-\mu^2)^{m/2}$. 
Thus $P_{l,l'}^{L,L',m}(k)$ gets multiplied with powers of $(1-\mu^2)^{m}$ or $\mu (1-\mu^2)^{m}$ to 
$\mu^{l+l'-2m}(1-\mu^2)^{m}$, so the highest order is $\mu^{l+l'}$. In addition we have $\mu^{L+L'}$ dependence in 
equation \ref{eq:p_ss}, so the lowest contribution in 
powers of $\mu$ to 
$P^{ss}(k)$ is $\mu^{L+L'}$ if $L+L'$ is even or $\mu^{L+L'+1}$ if $L+L'$ is odd, 
and the highest is $\mu^{2(L+L')}$. Thus for $P_{00}(\vk)$ the only angular term is isotropic, 
for $P_{01}(\vk)$ the only angular term is $\mu^2$, $P_{11}(\vk)$ and $P_{02}(\vk)$ contain both $\mu^2$ and $\mu^4$ etc. 
Note that only even powers of $\mu$ enter in the final expression, as required by the symmetry. 
We now proceed to look in more detail at the lowest order terms. 

\subsection{$P_{00}(\vk)$: the isotropic term}

At the lowest order in the expansion we have correlation of real space density $T_{||}^0=\delta(\vk)$ with itself. 
Density is a scalar of rank 0, $P_0(\mu)=1$. The power spectrum is isotropic
$P_{00}(\vk)=P_{00}^{000}(k)$. 
This term is just the real space power spectrum and of 
course does not have any $\mu$ dependence since it is independent of redshift space distortions. 
For small values of $\mu$ this term always dominates, and in the limit $\mu=0$ the transverse 
power spectrum becomes the real power spectrum $P_{00}(k)$. The real space power spectrum agrees with the
linear one on large scales, $P_{00}(k)=P_{\rm lin}(k)$, slightly dips below the linear one around $k \sim 0.1h/Mpc$, 
while on even smaller scales the nonlinear corrections cause it to increase over the linear one. 

\subsection{$P_{01}(\vk)$}

At the next order in our expansion (not in perturbation theory PT, see below) we have correlations between the density 
$T_{||}^0(\vk) =\delta(\vk)$ and radial component of momentum density 
$T_{||}^1(\vk)=[(1+\delta)u_{||}](\vk)$. 
Momentum density can be decomposed into a scalar ($m=0$) $T_1^{1,0}$
and two vector ($m=\pm 1$) components $T_1^{1,\pm 1}$, but only the scalar part correlates with the density $T_0^{0,0}$, 
which is a scalar. 
Thus the only contribution comes from $P^{0,1,0}_{0,1}(k) \propto \langle T_0^{0,0}(\vk)(T_1^{1,0}(\vk))^* \rangle$, 
\be
P_{01}(\vk)=P^{0,1,0}_{0,1}(k)\mu,
\ee
where we used $P_1^0(\mu)=\mu$. 

The scalar mode of momentum can be obtained from the divergence of momentum 
and related to $\dot{\delta}$ using the continuity equation,
which in terms of our quantities is
\begin{equation}
\dot{T}_0^{0,0}+ikT_1^{1,0}=0.
\label{ce}
\end{equation}
This is an exact relation (for conserved quantities), in the sense that the 
vector part of momentum does 
not contribute to it, since 
it vanishes upon taking the divergence (i.e., vector components are orthogonal to $\vk$ and the dot product is zero). 

From this we get that
\begin{equation}
P_{01}(\vk)=-ik^{-1}\mu P_{\delta,\dot{\delta}}(k)=-{i\mu \over 2k} {dP_{00}(k) \over dt}.
\end{equation}
The total contribution from this term to $P^{ss}(k)$ is 
\be
P^{ss}_{01}(\vk)=\mu^2 \HH^{-1} {dP_{00}(k) \over d\tau}=\mu^2 {d P_{00}(k) \over d\ln a}.
\ee
This is an exact relation for dark matter, valid also in the nonlinear regime. 
It shows that this term can be obtained directly from the redshift evolution of 
the dark matter power spectrum $P_{00}(k)$. On large scales it agrees with the linear theory predictions. 
If we write 
$P_{00}(k)=D(a)^2P_{\rm lin}(k)$, with $D(a)$ the linear growth rate and $f=d\ln D/d\ln a$, then 
we find $P^{ss}_{01}(\vk)=2f\mu^2P_{\rm lin}(k)$. 
We thus see that this order is of the same order in PT as the first order $P_{00}(k)$, the well known 
Kaiser result \cite{1987MNRAS.227....1K}.
On smaller scales we expect the term to deviate from
the linear one, just as for $P_{00}(k)$. 

\subsection{$P_{11}(\vk)$}

The next term is the correlation of the momentum density $T_{||}^1(\vk)$ with itself.
In this case the scalar ($m=0$) $T_1^{1,0}(k)$ 
correlates with itself and the vector ($m=\pm 1$) components $T_1^{1,\pm 1}(k)$ also 
correlate with itself, so both components of 
momentum contribute, 
\be
P_{11}(\vk)=P^{1,1,0}_{1,1}(k)[P_{1}^{0}(\mu)]^2+P_{1,1}^{1,1,1}(k)[P_{1}^{1}(\mu)]^2.
\ee
In terms of the contribution to the redshift space power spectrum this gives 
\be
P^{ss}_{11}(\vk)=\HH^{-2}k^2\mu^2[P^{1,1,0}_{1,1}(k)\mu^2+P^{1,1,1}_{1,1}(k)(1-\mu^2)]. 
\label{pss11}
\ee
The scalar part of the momentum is the one that contributes to the continuity equation \ref{ce}. 
In linear perturbation theory only the scalar contribution is non-zero and $P^{1,1,0}_{1,1}(k)=f^2P_{\rm lin}(k)$. 
This term is also of linear order and collecting all terms at this order we obtain the usual expression \cite{1987MNRAS.227....1K}
\be
P^{ss}_{\rm lin}(\vk)=(1+f\mu^2)^2P_{\rm lin}(k). 
\ee

However, we see from the above that there will be another contribution to both $\mu^2$ and $\mu^4$ 
terms from the vector part of momentum correlator $P_{1,1}^{1,1,1}(k) \propto \langle |T_1^{11}(k)|^2\rangle$, 
which comes in at the second order in power spectrum. 
This vector part is often called the vorticity part of the momentum. 
In general this term is non-zero because vorticity of momentum does not vanish, 
even if vorticity of velocity vanishes for a single streamed fluid  
\cite{2002PhR...367....1B}.
As seen from equation \ref{pss11} this term always {\it adds} power to $\mu^2$ term and subtracts it in $\mu^4$ term (but is combined with 
a positive contribution from the scalar part in $\mu^4$ term). 

\subsection{$P_{02}(\vk)$}

At orders higher than $P_{11}(\vk)$ we no longer have any linear contributions, hence these terms are usually 
not of interest for extracting the cosmological information. 
However, these terms, including what is sometimes called the Fingers-of-God 
(FoG) effect, are known to be important on 
fairly large scales. 
Here we will limit the discussion to some general statements of their $k$ and 
$\mu$ dependence, 
leaving their more precise calculations to future work. 

There are two different terms that contribute to this term,
\be
P_{02}(\vk)=P^{0,2,0}_{0,0}(k)[P_{0}^{0}(\mu)]^2+P^{0,2,0}_{0,2}(k)P_0^0(\mu)P_{2}^{0}(\mu).
\ee
In terms of the contribution to the redshift space power spectrum this gives
\be
P^{ss}_{02}(\vk)=-\left({k \mu \over \HH}\right)^2\left[P^{0,2,0}_{0,0}(k)+{1 \over 2}P^{0,2,0}_{0,2}(k)(3\mu^2-1)\right]. 
\ee
The first term is the correlation between the isotropic part of the mass weighted square of velocity, i.e. the energy density,
$T_0^{2,0}=(1+\delta)u^2$ and the density field $T_0^{0,0}=\delta$. The second term comes from the scalar part 
of the anisotropic stress $T_2^{2,0}$ correlated with the density $T_0^{0,0}=\delta$. 

On physical grounds we expect the first term 
to be large in systems with a large rms velocity resulting in a term scaling as 
$P^{0,2,0}_{0,0}(k) \sim P_{00}(k)\sigma^2$, where $\sigma^2$ has units of velocity squared, but is not 
simply the volume averaged velocity squared (see below). The contribution of this term to $P^{ss}$
goes as $-(k\mu/\HH)^2\sigma^2 P_{00}(k)$, i.e. it is a damping term suppressing the linear power spectrum, 
with the effect increasing towards higher $k$ (smaller scales). 
This is the lowest order FoG term, which we see contributes as $(k\mu)^2$ dependence and so affects the $\mu^2$ term.  
It is a damping term that is always negative, while the corresponding $\mu^2$ term from $P_{11}$ always adds power. 
The scalar anisotropic stress-density correlator $P_{0,2}^{0,2,0}(k)$ 
also contributes to $\mu^2$ angular term, as well as to $\mu^4$ angular term, 
and is formally of the same order in perturbation theory as 
$P^{0,2,0}_{0,0}(k)$, 
but is likely to be smaller
on physical grounds that velocity dispersion in virialized objects 
is isotropic and hence has a small anisotropic stress. 

\subsection{$P_{03}(\vk)$, $P_{12}(\vk)$, $P_{04}(\vk)$, $P_{13}(\vk)$ 
and $P_{22}(\vk)$}

Since the lowest order in $\mu$ is $(L+L')$ or $L+L'+1$, 
the terms of order higher than $P_{02}(\vk)$ do not  contribute to $\mu^2$ 
term. 
At the next order in $\mu$ we have terms of order $\mu^4$. At this order there
are 
7 terms that contribute. $P_{11}(\vk)$ and $P_{02}(\vk)$ we already discussed: 
while $P_{11}(\vk)$ has a linear order term and 
is expected to dominate on large scales, $P_{02}(\vk)$ is second order in power
spectrum. They both come with a prefactor of $k^2$. 
At the next order we have $P_{03}(\vk)$ and $P_{12}(\vk)$, both second order in power spectrum 
and each multiplied by $k^3$, followed by $P_{13}(\vk)$ and $P_{22}(\vk)$, also second order in power spectrum, but 
each multiplied by $k^4$, and by $P_{04}(\vk)$, third order in power spectrum. 
All of these terms also
contribute to terms of higher order in $\mu^{2j}$, up to 
$\mu^6$ or $\mu^8$ for these terms. 

One can see from this discussion that the angular structure of higher order 
terms is considerably more 
complex than that of lower order terms and that all the terms even in powers of
$\mu$ are being generated by 
RSD. However, there is a connection between the angular order, powers of $k$ 
and lowest order in perturbation 
theory, such that only the low powers of $k$ and low lowest order of PT 
contribute to the lowest orders in $\mu$. 
Thus, at low values of $k\mu$, the series is convergent. To make these 
statements more quantitative 
a numerical or perturbative analysis is required, which will be presented 
elsewhere \cite{Okumura2011}.

\subsection{Shot noise and connections between the correlators}

The correlators at the same order in powers of velocity, i.e. equal $L+L'$, 
contain nontrivial cancellations among them. 
To see this assume velocity is constant over a region of space 
$r \sim k_0^{-1}$. 
For example, large scale bulk flows lead to correlated velocities on small 
scales, giving rise to 
nearly equal velocities between nearby particles.  On scales smaller than this 
velocity coherence scale, $k>k_0$, we  
can pull out these constant velocity terms from the correlators, to obtain  
\be
P_{LL'}(k)\delta_D(\vk-\vk')=\langle [(1+\delta)u_{\parallel}^L](\vk)[(1+\delta)
u_{\parallel}^L]^*(\vk')\rangle \sim P_{00}(k) \langle u_{\parallel}^{L+L'} 
\rangle \delta_D(\vk-\vk')
,
\ee
where $\langle u_{\parallel}^{L+L'} \rangle$ is just a number corresponding to 
the spatial average of this term. 
So these terms are all equal as long as $L+L'$ is the same. 

These terms enter into the sum in equation \ref{eq:p_ss0} with different 
prefactors and opposite signs, leading to 
cancellations between them. The lowest order example is that of $P_{11}$ and 
$P_{02}$, which enter with equal prefactors but 
opposite signs, canceling any such contributions from each other. 
This is not surprising: bulk flows lead to ``rigid body'' displacements 
of particles but do not contribute to FoG effects, so their contribution to 
$P_{02}$ must be canceled. 
As a result, only velocity dispersion type contributions lead to FoG effects. 

In the extreme case this argument can be applied to the shot noise
for these correlators, which is the 
contribution to the power spectrum caused by discreteness
of tracers. It is well known that the shot noise of a density field sampled by 
tracers of number density $\bar{n}$ is 
given by $P_{00}(k)=\bar{n}^{-1}$. Analogous calculation for the moments gives 
\be
P_{LL'}(k)=\bar{n}^{-1}\langle u_{\parallel}^{L+L'}\rangle.
\ee
This expression is exact, since by definition a discrete tracer population only has a single value of velocity at any given position, 
so $\langle u_{\parallel}^{L+L'} \rangle$ will be the same for any pair of $L,L'$ such that $L+L'$ is the same. 
These shot noise terms can be large if the tracer is sparse, i.e. if $\bar{n}$ is small. 
However, the argument above shows that these terms enter with opposiste signs in the final result and so these shot 
noise contributions cancel in the total sum of \ref{eq:p_ss0}. This is expected: the only shot noise contribution 
to the total RSD power spectrum $P^{ss}(k)$ should be $\bar{n}^{-1}$. 
These examples show that these velocity moments are connected, and 
it is more natural to consider
them together, such as $P_{11}(k)-P_{02}(k)$, where the shot noise and the bulk flow terms cancel out. 

\subsection{Relation to Legendre moments}

In RSD analyses it is customary to integrate $P^{ss}(\vk)$ over the lowest order Legendre polynomials to obtain 
moments $P^{ss}_l(k)$, 
\be
P^{ss}_{l}(k)=(2l+1)\int_0^1 P^{ss}(\vk)P_l(\mu)d\mu,
\ee
where $P_l(\mu)$ are the ordinary Legendre polynomials, $P_0(\mu)=1$, 
$P_2(\mu)=(3\mu^2-1)/2$ and $P_4(\mu)=(35\mu^4-30\mu^2+3)/8$.
Only the lowest 3 orders contain contributions from linear terms, so the analysis is usually limited to $l=0,2,4$. 
The 0 moment is just the spherical average of the power spectrum in redshift space. 
The advantage of this expansion is that in a typical survey the moments are uncorrelated on scales small compared to 
the size of the survey. 

Moments in even $l$ can be viewed as an alternative way to expand in terms of even powers of $\mu$. However, 
the expansion given in equation \ref{pll} is {\it not} an expansion in Legendre polynomials, since it contains 
products of associated Legendre polynomials (including squares of ordinary Legendre polynomials of both even and odd
orders). Hence there is no orthogonality between the moments of distribution function and Legendre moments $P^{ss}_{l}(k)$. So 
if we expand the angular dependence of any given term $P^{ss}_{LL'}(k)$ into Legendre polynomials, we will 
generate all orders up to $l=2(L+L')$. This means for example that all terms
will contribute to the monopole $l=0$, all except $P^{ss}_{00}(k)$ to quadrupole 
$l=2$ and all but $P^{ss}_{00}(k)$ and $P^{ss}_{10}(k)$ to hexadecupole $l=4$. 
As a result we always have an infinite number of terms $P_{LL'}(\vk)$ contributing to any given
Legendre series term $P^{ss}_l(k)$, with 
higher and higher powers of $k$. This expansion is thus considerably more 
complex than the expansion in powers of $\mu^{2j}$, 
which has a finite number of terms for any given value of $j$. 

The discussion above suggests it may be more beneficial to fit for powers of $\mu^{2j}$ rather than 
for Legendre moments, for example by fitting for $\mu^0$, $\mu^2$ and $\mu^4$ terms, which contain 
linear order contributions, together with higher order terms $\mu^6$, $\mu^8$ etc., which we do not care for and 
can marginalize over in the end. 
However, Legedre moments are uncorrelated while powers of $\mu^{2j}$ are strongly 
correlated, so a marginalization over higher order terms will lead to a large increase in errors  for higher $k$, given that these
terms become very large at high $k$ compared to lowe order terms. So 
this can only work if sufficiently strong priors are adopted for higher order terms $\mu^6$, $\mu^8$ etc. Such priors could 
come from simulations extracting individual higher order terms or from a parametrized model. This is pursued further in \cite{Okumura2011}.

\subsection{Applications to galaxies and issues of bias} 

The relation to other tracers such as galaxies is a rich subject worth exploring further with this method. 
In this paper we focus primarily on the dark matter, but all the derivations remain unchanged if the dark matter 
particles are replaced with some other tracers, such as galaxies or halos. 
In large scale structure we usually define bias as the ratio of galaxy power spectrum (shot noise subtracted) to 
matter power spectrum, 
$b^2(k)=P_{00}^{gg}(k)/P_{00}^{mm}(k)$. 
We can generalize the concept of bias to 
\be
b_{LL'}(\vk)={P_{LL'}^{gg}(\vk) \over P_{LL'}^{mm}(\vk)}, 
\label{bll}
\ee
where $P_{LL'}^{gg}(\vk)$ is galaxy correlator and $P_{LL'}^{mm}(\vk)$ is the corresponding dark matter term. 
In linear theory we have $b_{00}=b_1^2$, $b_{01}=b_1$ and $b_{11}=1$, independent of scale or angle, where 
linear bias is defined as $\delta_g=b_1\delta_m$. 
Two ways to extract cosmological information from RSD are either by combining $P_{00}$ and $P_{01}$ to eliminate $b_1$, or 
to use $P_{11}$ directly.  

Before discussing RSD further it is useful to draw a comparison to weak lensing. In case of weak lensing 
we can measure both projected dark matter density or galaxy density, so we can perform 
a joint correlation analysis of galaxy clustering and weak lensing, where the 
galaxy auto-correlation is proportional to $b^2$ times matter correlations, cross-correlation between galaxies and 
weak lensing signal around them induced by the dark matter is proportional to $b$ (the so called galaxy-galaxy lensing), 
while the weak lensing auto-correlation is independent of bias. 
Two ways to extract the signal are either using just shear-shear correlations tracing matter-matter correlations, 
or combining galaxy auto-correlation with galaxy-galaxy lensing to eliminate bias. 
This latter has higher signal to noise but is complicated due to the fact that bias is 
scale independent and the scale dependence depends on the 
galaxy properties \cite{2010PhRvD..81f3531B}. To understand when this happens it is useful to expand galaxy density perturbation 
to second order in matter density, $\delta_g=b_1\delta_m+b_2\delta_m^2$. 
The second order terms will become important when they cannot be neglected against the first order terms,
so the expansion parameter is $(b_2/b_1)\delta_m$. Since $\delta_m^{\rm rms}$ increases on small scales this scale 
dependent bias increases towards small scales. Typically we have $|b_2/b_1|<0.4$ \cite{2010PhRvD..81f3531B} 
and the corrections become 
important at $k \sim 0.1h/Mpc$, where $\delta_m^{\rm rms} \sim 0.5$.  

Returning back to RSD, our formalism is directly applicable to galaxies, except that 
all the velocity moments are mass weighted for the dark matter, $T_{\parallel}^{L,m}=
(1+\delta_m)u_{\parallel}^L$, and number density weighted for the galaxies, $T_{\parallel}^{L,g}= 
(1+\delta_g)u_{\parallel}^L$. 
What this shows is that if the density distribution of galaxies differs from that of the
dark matter then all the correlators of velocity moments will differ from each other, even those that appear independent of bias, 
such as $P_{11}$. 
In reality thus the predictions of linear bias model will be modified, 
because even if galaxies are faithful tracers of the dark matter velocities at a given position, the weighting of
the velocity moments differs: in one case they are weighted by the dark matter mass, in the other by the number of 
galaxies, and the two 
differ in their spatial distribution. This will result in scale dependence of the higher order 
bias terms $b_{LL'}$, just like it does for the $b_{00}$ itself \cite{2010PhRvD..82d3515H}. 

To quantify this further, for the lowest order momentum density term and for linear bias
we must compare correlation of $(1+b_1\delta_m)u_{\parallel}$ with itself  to give $P_{11}^{gg}(\vk)$ 
or with $b_1\delta_m$  to give $P_{01}^{gg}(\vk)$. The auto-correlation will give the result that agrees with the dark matter 
only for $b_1=1$, or if $b_1\delta_m \ll 1$. In the same limit the cross-correlation will give linear bias $b_1$. 
So momentum density becomes velocity in the limit $b_1\delta_m \ll 1$,
while requiring it to be scale independent relative to dark matter  
requires something like $(b_1-1)\delta_m \ll 1$, which for typical LRG galaxies ($b_1 \sim 2$) 
is in fact a more stringent requirement than that of a scale independent bias condition 
discussed above 
$(b_2/b_1$ versus $b_1$). This  suggests 
that the scale dependence of the momentum density bias terms $b_{01}$ and $b_{11}$ 
defined in equation \ref{bll} extends to larger scales
than scale dependent bias of density $b_{00}$. 

The conclusion from this discussion is that the scale dependence of bias terms 
involving momentum density is a real concern in RSD and likely 
extends to relatively large scales ($k<0.1h/Mpc$). 
In terms 
of the angular decomposition in powers of $\mu$, 
the discussion of the scale dependent bias of RSD can be divided into a $\mu^2$ term, which depends 
entirely on $b_{01}$ in $P_{01}$, and the $\mu^4$ term, for which 
the scale dependence of $b_{11}$ term above is applicable,
since that is the term that does not vanish on large scales in linear theory. 
In this sense RSD analysis is not the equivalent of a joint galaxy-weak lensing analysis, 
since weak lensing auto-correlation truly traces the dark matter directly, 
while in RSD 
this limit is achieved only on relatively large scales where 
$\delta_g^{\rm rms} \ll 1$. 

The discussion so far completely ignored FoG effects: for $\mu^2$ term 
these are encoded in $P_{02}$ and in vector part of $P_{11}$, which  
unlike $P_{02}$ adds power rather than removes it, and these terms 
have their own physical interpretation and scale dependence 
unrelated to the scale dependent bias discussion above. 
While they can partially cancel the effects discussed above they are unlikely to 
achieve this exactly. 
In most of the literature so far only the scale dependence induced by FoG effects was discussed 
(although see \cite{2004PhRvD..70h3007S,2011arXiv1105.4165R}). 
The simple linear bias model predicts that FoG effects scale with bias squared: the leading order term scales as $b_1^2$ 
both in $P_{02}^{gg} \propto 
\langle [b_1\delta v_{\parallel}^2](\vk)b_1\delta(-\vk)$ and in $P_{11}^{gg}\propto [b_1\delta v_{\parallel}]^2$. 
If we write $P_{02}^{gg}(k)-P_{11}^{gg}(k)=P_{00}^{gg}(k)\sigma^2$, then $\sigma$ is independent of bias, 
since $P_{00}^{gg}(k) \propto b_1^2P_{00}^{mm}(k)$. 

\section{Discussion}

In this paper we present a distribution function approach to redshift space distortions. 
We show that the redshift space density can be expressed in terms of a
sum over velocity moments and the redshift 
space power spectrum can be expressed in terms of correlators between the Fourier 
components of these moments. 
These moments are simple objects to calculate in any system: they are calculated by simply taking appropriate powers 
of radial velocity and summing over all particles. The lowest order moments are density, momentum density, stress 
energy density etc. 

We have decomposed the moments into helicity 
eigenstates based on their transformation properties under rotation around the direction of the Fourier mode, 
a generalization of SVT decomposition in cosmological perturbation theory. 
We use rotational invariance to derive all of the allowed correlator terms, showing that only terms with 
the same helicity can contribute to the correlators. 
The moments of distribution function are complicated objects with many terms allowed by symmetries, specially at higher order, 
leading to a complicated angular and scale dependence, 
suggesting that treatments of RSD cannot be fully successful with simple ansatzes, such as the popular FoG velocity
dispersion model with one free parameter \cite{2011MNRAS.410.2081J,2011arXiv1103.3614T}. 

Despite the complexity of the general RSD description some general statements can be made. 
The lower order terms generally only 
contribute to low orders of expansion in $\mu^2$, where $\mu$ is the angle between the Fourier mode and the line of sight. 
As an example, we have shown that only the scalar part of the momentum density correlates with the density, and 
this term can be written in terms of a time derivative of the power spectrum. This term only contributes to $\mu^2$
angular dependence and contains a linear order term. But 
there is also the vector part of the momentum density-momentum density correlation, 
the (scalar) energy density-density correlation, and the scalar part of anisotropic stress density-density correlation, 
all of which also contribute to the $\mu^2$ term. 
They are all nonlinear and cannot dominate on very large scales, but likely 
dominate on small scales. 
The energy density-density correlation term is the term most closely related to the 
FoG velocity dispersion effect and is always negative, suppressing the power, 
but the other terms are formally of the same order in perturbation theory. 
We have shown that the vorticity part of momentum always adds to the RSD power of $\mu^2$ term, and hence acts in the opposite direction to 
the FoG term. Our analysis cannot address which term has a larger amplitude, but it would be interesting to see 
if there are any systems where the terms that add power dominate over those that suppress it. 
The next angular term has $\mu^4$ dependence and we identified 7 terms that contribute to it, of which one, 
scalar part of $P_{11}$, contains a linear contribution that does not vanish on large scales. 

The fact that there are a finite number of velocity moment terms at each order of $\mu^{2j}$ expansion should be contrasted to
the popular Legendre multipoles expansion (monopole, quadrupole and hexadecupole contain cosmological information),
which receive contributions from all orders in moments of distribution function. 
This suggests that a better behaved analysis may be possible if instead of a multipole analysis the analysis is performed in 
terms of a $\mu^{2j}$ expansion, with the lowest 3 orders containing cosmological information and the rest treated as nuisance 
parameters. 

It is important to emphasize that these moments are mass weighted quantities, and 
no volume averaged quantities ever enter into our expressions. 
This relates to one of the long standing issues in the treatment of RSD: many of the past treatments \cite{2004ApJ...606..702T,2011MNRAS.410.2081J,2011arXiv1103.3614T} have 
assumed that RSD trace correlations between velocities and dark matter and that the FoG effects multiply these 
density-velocity and velocity-velocity correlations, where FoG quantities are also
defined as volume weighted quantities such as velocity dispersion $\sigma^2=\langle u^2 \rangle$. 
But these volume weighted quantities 
are not well defined, specially for sparse biased systems such as galaxies or clusters. 
For a biased tracer with $b>1$ one finds that voids with no tracers in them are enlarged, 
since, for $\delta<0$, $1+b\delta$ is closer to 0 than $1+\delta$. 
This has forced some workers to use the dark matter velocity field instead, 
with unpredictable results \cite{2011arXiv1103.3614T}. 
Our expansion shows that it is more natural to define RSD in terms of mass or number weighted quantities, such as momentum density or 
energy density, the former replacing velocity and the latter replacing velocity dispersion. 
Mass and number weighted moments such as momentum or energy density
are well defined even in voids (where they are simply zero). In this paper we show that there is a consistent 
expansion using mass weighted moments, and that the expansion is convergent on large scales. 

The fact that all RSD quantities are density weighted also suggests that RSD effects will differ 
if the galaxy number density distribution differs from mass density distribution. We have shown that even a linear 
bias model induces scale dependent bias of the momentum density correlators, and that this scale dependence is 
likely to show up on relatively large scales, $k<0.1{\rm h/Mpc}$. 
The success of RSD in extracting cosmological information depends entirely on our ability to model 
these various bias terms and relate them to each other. Similarly, 
the success of the approach presented here in modeling RSD depends on our ability to extract these moments from simulations and 
data and on our ability to model them with analytic models, such as perturbation theory. 
Providing physical interpretation of the terms, as done here, could enable one to develop more effective modeling, 
or provide a better physical understanding of limitations of RSD in extracting cosmological information.
For example, it is relatively straight-forward to include the bias induced scale dependence
effect at the lowest order of PT and we will present the results elsewhere \cite{Vlah2011}. 
In this paper we have focused on theory, conceptual issues and general symmetries, while applications to simulations 
and perturbation theory will be presented in 
upcoming work \cite{Okumura2011,Vlah2011}. 

\begin{acknowledgments}
We thank Teppei Okumura and Zvonimir Vlah for helpful discussions. 
This work is supported by the DOE, the Swiss National Foundation under contract 200021-116696/1 and WCU grant R32-10130. 
\end{acknowledgments}

\bibliography{cosmo,cosmo_preprints}
\bibliographystyle{revtex}

\end{document}